%% file: ms_nat.tex
\documentclass{natureWithFig}	

\usepackage{epsfig}
\usepackage{comment}
\usepackage{url}
\usepackage{amssymb}	

\newcommand{\arcsec}{\mbox{$''$}}
\newcommand{\arcmin}{\mbox{$'$}}

\newcommand{\citet}[1]{ref.~\citen{#1}}
\newcommand{\Citet}[1]{Ref.~\citen{#1}}
\newcommand{\citep}{\cite}

\newlength{\propwidth}
\newcommand{\flash}{GN-z11-flash}
\newcommand{\bree}{Breeze-M debris}

\input{abbreviations}

\title{GN-z11-flash was a signal from a man-made satellite not a gamma-ray burst at redshift 11
}

\author{Micha{\l}~J.~Micha{\l}owski$^{1}$, Krzysztof Kami{\'n}ski$^1$, Monika K.~Kami{\'n}ska$^1$ \& Edwin Wnuk$^1$
}

\begin{document}

\maketitle

\begin{affiliations}
 \item Astronomical Observatory Institute, Faculty of Physics, Adam Mickiewicz University, ul.~S{\l}oneczna 36, 60-286 Pozna{\'n}, Poland, {\tt michal.michalowski@amu.edu.pl}, {\tt chrisk@amu.edu.pl}
\label{inst:roe} 
\end{affiliations}


\begin{abstract}
Long gamma-ray bursts (GRB), explosions of very massive stars, provide crucial information on stellar and galaxy evolution, even at redshifts $z\sim8-9.5$, when the Universe was only 500--600 million years old\cite{salvaterra09,tanvir09,cucchiara11}.
Recently, during observations\cite{jiang20b} of a galaxy at a redshift of $z\sim11$ (400 million years after the Big Bang)\cite{oesch14,oesch16}, a bright signal, named {\flash}, shorter than 245\,s was detected and interpreted as an ultraviolet flash associated with a GRB in this galaxy\cite{jiang20a}, or a shock-breakout in a Population III supernova\cite{padmanabhan21}. Its resulting luminosity would be consistent with that of other GRBs\cite{kann20}, but a discussion based on probability arguments started on whether this is instead a signal from a man-made satellite or a Solar System object\cite{steinhardt21,jiang21,nir21}. 
Here we show a conclusive association of {\flash} with Breeze-M upper stage of a Russian Proton rocket on a highly elliptical orbit.
This rules out {\flash} as the most distant GRB ever detected. It also implies that monitoring of a larger sample of very high redshift galaxies is needed to detect such distant GRBs. 
This also highlights the importance of a complete database of Earth satellites and debris, which can allow proper interpretation of astronomical observations.

\end{abstract}

We searched Space-Track, the largest publicly available database of Earth satellites and space debris (\url{www.space-track.org}) for an object close to the position of {\flash} at the time of observations. We found the Breeze-M space debris (North American Aerospace Defense Command (NORAD) number 40386; Committee on Space Research (COSPAR) number 2015-005C), with a trajectory consistent with the position of GN-z11 at the time of the flash. It has a highly elliptical orbit with the semi-major axis of 13,946\,km, the perigee distance to the Earth's surface of 394.8\,km and the apogee distance of 14,757.2\,km. Its orbit is inclined by 50.7$^\circ$ to the Earth's equator and the orbital period is 273.2 minutes.

Using orbital elements from the database and our Satellite Trail Predictor software package (see Methods), we found that the closest angular separation between the satellite and the position of GN-z11 was $(18\pm30)${\arcsec}, consistent with zero (the offset from the edge of the slit used for the flash detection was $(7\pm30)${\arcsec}). This was at the UT time of 2017-04-07 08:07:50.80, within the integration of the frame with the flash detection, which started at 08:07:19.86 and lasted 179 sec\cite{jiang20a}.
{\bree} was moving at a position angle of 104.24$^\circ$ with the angular velocity of
35.9{\arcsec} per sec. At that time its distance from the Earth's surface was 13\,758\,km, whereas the distance to the observer in Hawai`i was 15\,186\,km. The satellite was also outside the Earth's shadow.

\Citet{jiang20a} presented several arguments against {\flash} being a signal from a satellite. We demonstrate that none of these concerns applies to {\bree}, because it falls in the narrow regions of the parameter space not ruled out by the authors.
First, \citet{jiang20a} estimated that a high-Earth orbit satellite would need to have an orbit inclined by more than $38^\circ$ to the Earth's equator.
Indeed, the inclination of {\bree} is 50.7$^\circ$. \Citet{jiang20a} stated that ``the compact spectral shape already suggests that the trajectory of any hypothetical moving object must be nearly perpendicular to the slit, or else the spectrum would be broader.'' This is indeed the case for {\bree}, as the angle between the slit and the satellite's trajectory is 118.74$^\circ$, so an extended spectrum is not expected above the measured width in the spatial direction, which is 20--25\% broader than the point spread function (Extended Data Fig. 1 of \citet{jiang20a}).

Second, the estimated $V$-band magnitude of {\flash} of 19.2\,mag was used to imply the actual visual magnitude of $5.5\,$mag\cite{jiang20a}. However, this calculation assumed an angular velocity of 20--30{\arcmin} per sec with respect to the observer, corresponding to a low-Earth orbit. This is 33--50 times faster than the actual velocity of {\bree}, so the satellite should be much fainter than 5.5\,mag.

Indeed, 
our archival observations carried out with the Roman Baranowski Telescope/Poznań Spectroscopic Telescope 2 (RBT/PST2; see Fig.~1 and Methods for the description) revealed the brightness of {\bree} with an infrared cut-off filter of $\sim9.2$\,mag at the distance of $\sim15,000$\,km from the observer (the distance to the observer during the flash detection). Taking into account the phase angle change, the estimated $V$ magnitude of {\bree|} at the time of the {\flash} observation is $\sim9.7$\,mag.
The detection of {\flash} was made with the Multi-Object Spectrometer For Infra-Red Exploration (MOSFIRE)\cite{mosfire} mounted at the Keck telescope, with the slit width of $0.9${\arcsec} and the position angle of 345.5$^\circ$. The angular speed of {\bree} implies that it was visible in the slit for around $0.029$ sec. Each MOSIFRE frame was integrated for 179 sec, so this means that the satellite appeared a factor of $179/0.029\sim6172$ times (or $\sim9.5$\,mag) fainter than it really was. This implies the expected magnitude of 19.2\,mag, close to the one measured from the MOSFIRE data. 
Given 
the uncertainties in the {\bree} brightness resulting from the irregular shape, rotation, albedo and changes of phase angle, the brightness recorded by MOSFIRE is consistent with our measurements for this satellite.
 
Third, the MOSFIRE observations were conducted simultaneously for 21 slits, so a moving object should produce a similar flash in one or more of other slits ``unless its trajectory was nearly perpendicular to the slit ($\sim 90^\circ\pm20^\circ)$''\cite{jiang20a}. Indeed, the trajectory of {\bree} was inclined by 118.74$^\circ$ with respect to the slit, just outside the interval quoted by \citet{jiang20a}. Hence, the fact that the flash was not visible through other slits was just because {\bree} was unlucky to miss all of them. The time gap between frames was 33 sec, during which the satellite moved 23{\arcmin}, i.e.~more than the field of view of MOSFIRE (6\arcmin). Hence, the satellite was not visible in the previous or next frame.

In order to demonstrate that {\bree} missed other slits, on Fig.~2 we show its trajectory together with the positions of the MOSFIRE slits obtained from the Keck Observatory Archive (KOA; Program ID: S324, PI: Kashikawa, frame name: MF.20170407.29239). 
We show the best orbit with the $1$, $2$, and $3\sigma$ confidence intervals (see Methods). In order to show systematic uncertainties we also show trajectories resulting from the orbital elements available for epochs between one week before and after the MOSFIRE observations. Trajectories crossing the slit corresponding to the {\flash} position and missing all other slits are consistent with the best orbit at a level below $1\sigma$. Moreover, if this satellite was not associated with {\flash}, then it would most probably be observed at another slit at this time, as there are very few trajectories within the confidence interval that do not cross any slit.

Hence, we conclude that {\flash} was the signal from {\bree}, which had a position consistent with GN-z11 at the time of observations and all other properties consistent with observational results. Hence, we confirm the hypothesis of \citet{steinhardt21} and \citet{nir21} that this was a man-made signal. This also highlights the importance of a complete database of Earth satellites and debris, which allows the interpretation of astronomical observations.

\section*{Methods}

\subsection{Orbit analysis software.}

The trajectory of the satellite was calculated using our Satellite Trail Predictor software package, which is based on the Simplified General Perturbation (SGP) orbital models\cite{Vallado2006}. This software has recently been developed by K.K.~and M.K.K.~specifically for the purpose of predicting the position, orientation and brightness of satellite trails in astronomical images and will soon be released as a publicly available on-line service (\url{www.astro.amu.edu.pl/STP/}).

In order to verify these orbit calculations we also used the following publicly available software packages: JPL Horizons\cite{jplhor1,jplhor2} (\url{ssd.jpl.nasa.gov/horizons.cgi}), OREKIT\cite{orekit} (\url{www.orekit.org}), and SkyField\cite{skyfield} (\url{rhodesmill.org/skyfield/}). The resulting orbits agree with our calculations within 0.1\,km (1.5\arcsec).

\subsection{RBT data.}

We used our archival data from the Roman Baranowski Telescope/Poznań Spectroscopic Telescope 2 (RBT/PST2; http://www.astro.amu.edu.pl/GATS), a 0.7m robotic telescope located at Winer Observatory,
USA. We observed {\bree} 16 times during 2020-05-16 to 2020-10-15 with an infrared cut-off filter and real-time telescope satellite tracking. We measured the magnitudes using elongated apertures for reference stars and nearly circular aperture for the satellite. The resulting brightness is shown on Fig.~1 as a function of the distance to the observer.

We also used these observations to estimate the uncertainty of our orbit calculations. For each of the 16 nights we used the Space-Track orbit with the epoch corresponding to the observation date in the same way as for the time of the flash observations. Then we compared these calculations to the astrometric positions measured from our RBT data. We measured the mean offset perpendicular to the line of sight of 1.7\,km with a standard deviation of 2.2\,km. At the distance of {\bree} from the Keck telescope at the time of the flash observations of 15,186\,km, this corresponds to the mean offset between the measured position and the position calculated from the orbit of  23{\arcsec} with a standard deviation of 30{\arcsec}. We adopt the latter value as the positional uncertainty. 

This uncertainty is consistent with that typically found for orbital models. Typical errors of the SGP orbital models are 1\,km and the orbital data errors perpendicular to the orbit are 1.1\,km for highly elliptical orbit satellites\cite{flohrer09}. This corresponds to a total error of the trajectory of 1.5 km, or 21{\arcsec} at the distance of 15\,186\,km.



\begin{addendum}
\item[Supplementary Information] is linked to the online version of the paper at www.nature.com/nature.
 \item
 M.J.M.~acknowledges the support of 
the National Science Centre, Poland through the SONATA BIS grant 2018/30/E/ST9/00208. We acknowledge the use of the Space-Track database, JPL Horizons, OREKIT, and SkyField packages.
This research has made use of the Keck Observatory Archive (KOA), which is operated by the W. M. Keck Observatory and the NASA Exoplanet Science Institute (NExScI), under contract with the National Aeronautics and Space Administration.

 \item[Author Contributions] 
 M.J.M.~designed this project and led the manuscript writing, K.K.~and M.K.K.~searched for the satellite, calculated its properties, planned and analysed the RBT observations, made the figures, and contributed to the manuscript writing. E.W.~also calculated the trajectory of the satellite.
 \item[Author information] 
 Reprints and permissions information is available at www.nature.com/reprints.
The authors declare no competing financial interests.
Correspondence and requests for materials should be addressed to M.J.M.~(email:  {\tt michal.michalowski@amu.edu.pl}, or K.K. (email: \\ {\tt chrisk@amu.edu.pl}).
\end{addendum}

\newpage

\clearpage

\begin{figure}
\begin{center}
\includegraphics[width=\textwidth,clip]{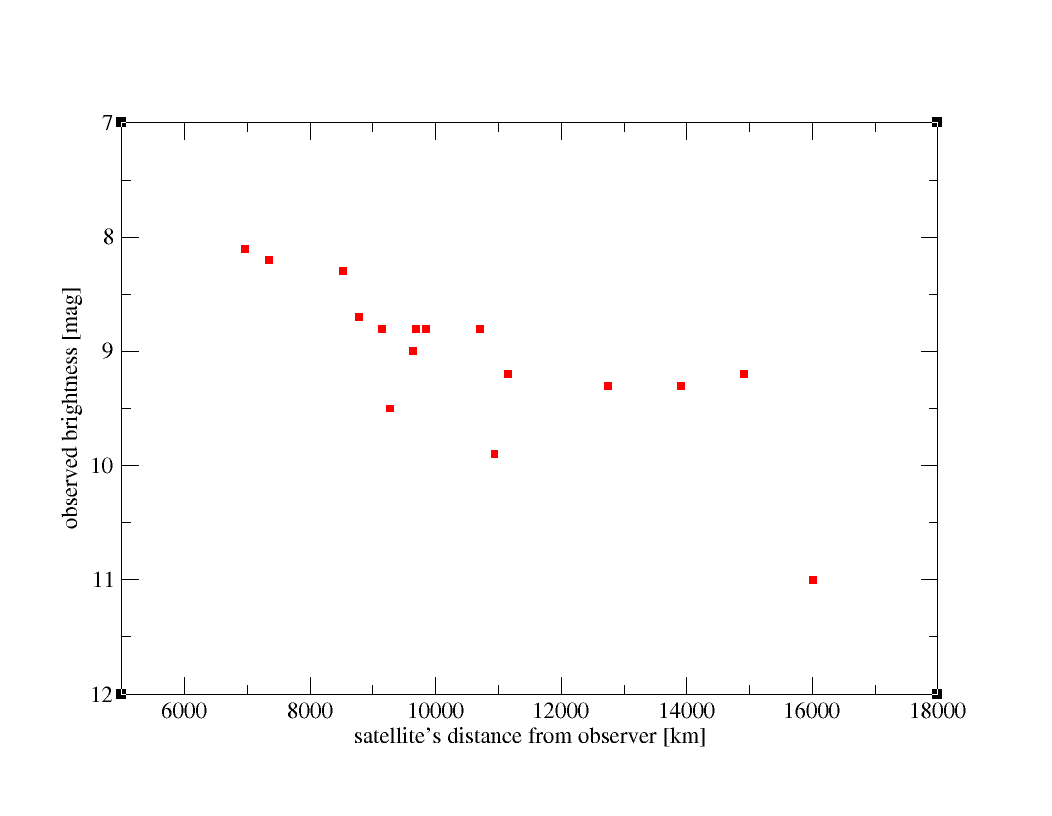} 
\end{center}
\end{figure}
\vspace{-1em}
{\bf Fig.~1.
Brightness  of {\bree} as a function of distance from the observer from our archival observations with the RBT/PST2 telescope with an infrared cut-off filter (see Methods). At the distance of around 15,000\,km, corresponding to the distance for the MOSIFRE observations, the brightness was $\sim9.2$\,mag, consistent with the MOSFIRE measurement taking into account phase angle differences and the angular speed of the satellite.}

\begin{figure}
\begin{center}
\includegraphics[width=\textwidth,clip]{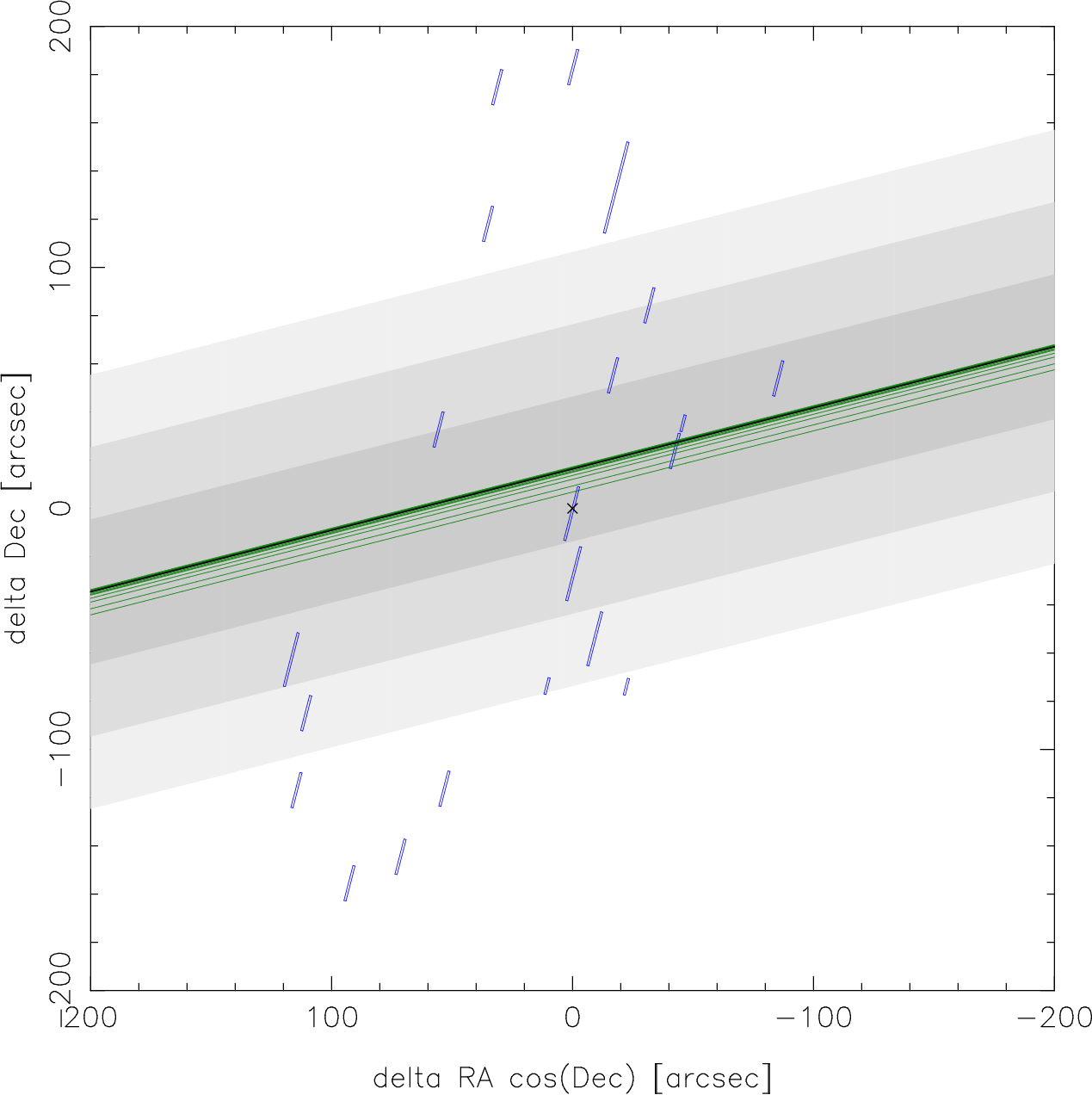} 
\end{center}
\end{figure}
\vspace{-1em}
{\bf Fig.~2. 
The trajectory of {\bree} in the field of view of MOSFIRE/Keck during the {\flash} detection. The trajectory for the orbit corresponding to the date of the observation is shown as a thick black line with $1$, $2$, and $3\sigma$ confidence intervals shown as shaded regions. The trajectories for orbit elements calculated for days up to one week before and after the MOSIFRE observations are shown as green lines. The blue rectangles correspond to the positions and sizes of the slits used during the MOSFIRE observations. The position of GN-z11 is shown as the black cross. Within $1\sigma$ of the best orbital model there are solutions crossing the slit corresponding to {\flash} and missing all other slits.}


\input{ms_nat.bbl}
\end{document}

%% file: abbreviations.tex
  \def\apj{ApJ}

 \def\nat{Nature}
